# Skyrmion annihilation in helimagnets and the critical condition of topological phase transitions with a soft mode

*Yangfan Hu**,

*Research Institute of Interdisciplinary Science & School of Materials Science and Engineering, Dongguan University of Technology, Dongguan, 523808, China.*

**Abstract**

Since the 1950s, topological solitons have been used to describe elementary particles[1-3] and particle-like field configurations[4-13] that appear in almost all branches of physics ranging from subatomic to cosmological scales[3,14-16]. In this context, formation and annihilation of particles corresponds to topological phase transitions, whose actual process and mechanism remain elusive. Here, we show that annihilation of an isolated skyrmion in helimagnets as the applied magnetic field increases initiates at its center point by a local magnetization reversal. At the critical field, a soft mode of the spin-wave excitation appears, which vanishes the local magnetization modulus as well as the soft-mode-modulated emergent elastic stiffness[17] at the center, which makes it an emergent magnetic monopole[6]. This softened point vanishes the energy barrier of reversing the local magnetization at the center, which inevitably annihilates the skyrmion. On this basis, we propose a general critical condition for topological phase transitions of field solutions with a soft mode: topological phase transitions initiate at those spatial points where the soft-mode-modulated emergent stiffness matrix losses its positive-definiteness.

**Main**

The concept of skyrmion is introduced by Tony Skyrme[18] initially in particle physics, and later widely studied in different physical systems such as string theory[15], Bose-Einstain condensates[19], 2-D electron gas[20], liquid crystals[21], chiral magnets[22, 23], and local photonic spin field[24]. Skyrmions or other emergent particles that appear in helimagnets or other condensed matter systems share similar mathematical structure with elementary particles, but have much lower excitation energy (some can exist as ground states[25]), larger size, and slower response time, for which they provide a better platform for us to understand the various basic

* Corresponding author: huyf@dgut.edu.cn

physical properties of elementary particles.

Beside the general scientific interest above, the formation and annihilation mechanism of an isolated magnetic skyrmion in chiral magnets, and various methods developed[6, 26-31] to precisely control its formation and annihilation is of great technical significance itself. Last decade has witnessed a strong revival of interest in magnetic skyrmions and skyrmionic structures, mainly because they show great prospective in realizing several next generation spintronics devices[32] such as skyrmion-based racetrack memories[33], neuromorphic computing[34, 35], and probabilistic computing[36]. In all these application scenarios, a key point is how to realize precise and reproducible formation, reading, and annihilation of a skyrmion[32].

We consider the three-dimensional magnetization vector field with changeable modulus in B20 chiral magnets, which permits a static soliton solution called an isolated magnetic skyrmion of the Bloch type[23, 25] (Fig. 1(a)). The free energy density of the system is given in a rescaled form[37] as

$$\phi = \sum_{i=1}^{3}\left(\frac{\partial \mathbf{m}}{\partial x_i}\right)^2 - 2\mathbf{b}\cdot\mathbf{m} + \mathbf{m}\cdot(\nabla\times\mathbf{m}) + \widetilde{T}\mathbf{m}^2 + \mathbf{m}^4, \qquad (1)$$

where the vector field $\mathbf{m}$ denotes the magnetization vector, $\mathbf{b}$ denotes the external magnetic field vector (here we assume $\mathbf{b} = [0, 0, b]^T$), and $\widetilde{T}$ is the rescaled temperature. We describe the problem in cylindrical coordinates $\mathbf{x} = [\rho cos\varphi, \rho sin\varphi, z]^T$, and the isolated skyrmion solution takes the form

$$\mathbf{m} = m(\rho)[-\sin\theta(\rho)\sin(\varphi+\psi(\rho)),\ \sin\theta(\rho)\cos(\varphi+\psi(\rho)),\ \cos\theta(\rho)]^T, \quad (2)$$

where $\psi(\rho) = 0$, and $m(\rho)$ and $\theta(\rho)$ are to be numerically determined by solving the Eular-Lagrangian equation with the boundary conditions $\theta(0) = \pi$, $\left(\frac{\partial m}{\partial \rho}\right)_{\rho=0} = 0$, and $\theta(R) = 0$, where $R$ is the radius of an isolated skyrmion (see Methods A for details). The profiles of $m(\rho)$ and $\theta(\rho)$ solved at different values of $b$ is shown in Fig. 1(b, c).

For the solution form given in eq. (2), as $b$ increases from zero, two subsequent topological phase transitions will occur, corresponding to two critical magnetic field: $b_s = 1.11$ and $b_a = 1.365$. At $b_s$, $R$ approaches infinity[17, 23]. This phenomenon can be regarded as the single particle version of the Kosterlitz-Thouless phase transition[38-39], so we call it a single-particle Kosterlitz-Thouless (SPKT) phase transition[17]. An appropriate choice of topological order parameter in this case is the averaged topological density $\rho_t = \frac{1}{4\pi^2 R^2}\int \mathbf{m}\cdot(\mathbf{m}_x\times\mathbf{m}_y)dS$. For finite $R$, $\rho_t$ is a finite real number, and for $R \to \infty$, $\rho_t \to 0$. The SPKT phase transition is a second-order topological phase transition, which can be explained by softening of the emergent elastic stiffness matrix according to the general theory of emergent elasticity[17]. For the free energy density given in eq. (1) and the soliton solution form given in eq. (2), we have four eigenvalues of the emergent stiffness matrix (see Methods B for details): $C^e_{\rho\rho} = 2[m'^2(\rho) + m^2(\rho)\theta'^2(\rho)]$, $C^e_{\varphi\varphi} = \frac{2m^2(\rho)\sin^2\theta(\rho)}{\rho^2}$, $C^e_3 = 4[m'^2(\rho) + m^2(\rho)\theta'^2(\rho)]$, and $C^e_4 = 4m^2(\rho)\frac{\sin^2\theta(\rho)}{\rho^2}$. Here, the parameter of interest is the homogeneous magnetic field $b$ applied along the $z$ axis, which corresponds to axially symmetric loading, and the only effective emergent stiffness coefficient is $C^e_{\rho\rho}$. As shown in Fig. 1(d, e), the SPKT phase transition is induced by softening of the emergent elastic stiffness component $C^e_{\rho\rho}(\rho = R)$ at $R \to \infty$.

The phase transition that occurs at $b_a = 1.365$ is a first-order topological phase transition, and it eventually leads to annihilation of the isolated skyrmion. Analysis of the emergent elasticity shows that at $b_a$ $C^e_{\rho\rho}(\rho)$ is positive for all $\rho$, although a sudden drop of $C^e_{\rho\rho}(\rho = 0)$ occurs near $b_a$ (Fig. 1(e)).

To understand why this phase transition occurs, we study the axially symmetric collective spin-wave dynamics of the system by solving the Landau-Lifshits (L-L) equation[40]

$$\dot{\mathbf{m}} = \gamma \mathbf{H}_{eff} \times \mathbf{m}, \quad (3)$$

where $\gamma$ denotes the gyromagnetic ratio, $\mathbf{H}_{eff} = \frac{1}{M_s}\frac{\delta\phi}{\delta\mathbf{m}}$ denotes the effective magnetic field and $M_s$ denotes the saturation magnetization. We consider the following form of magnetization distribution

$$\mathbf{m} = (m + m_v)[-\sin(\theta + \theta_v)\sin(\varphi + \psi_v), \sin(\theta + \theta_v e^{i\omega t})\cos(\varphi + \psi_v), \cos(\theta + \theta_v)]^T, \quad (4)$$

where $m_v = \widetilde{m}(\rho)e^{i\omega t}$, $\theta_v = \widetilde{\theta}(\rho)e^{i\omega t}$ and $\psi_v = \widetilde{\psi}(\rho)e^{i\omega t}$ denotes small amplitude vibration of $m$, $\theta$, and $\psi$ around their equilibrium state. By solving the linearized L-L equation for $m_v$, $\theta_v$ and $\psi_v$, we obtain the magnon spectrum for the isolated skyrmion state (see Methods C for details). In Fig. 2(a) we plot the variation of the rescaled magnon frequency $\widetilde{\omega}_i(b) = \omega_i(b)/\omega_i(b = 1.1)$, $(i = 0, 1, 2, \ldots, 9)$ with $b$ for the first ten spin-wave modes, where at $b = b_a$, $\widetilde{\omega}_0$ vanishes and become a soft mode. In Fig. 2(b-d) we show the vibration shape of the first ten spin-wave modes of $\widetilde{\theta}(\rho)$, $\widetilde{m}(\rho)$, and $\widetilde{\psi}(\rho)$.

The next question to ask is, why does the appearance of a soft mode lead to annihilation of the skyrmion, or in other words, how does this first-order topological phase transition occur. Soft mode corresponds to a collective dynamical pattern that losses its restoring stiffness, whose vibration magnitude increases with time. At finite temperature, the statistical average of this vibration magnitude will reach infinite, while concerning possible Gilbert damping of the system, its value will be finite but significant. As a result, to analyze the behavior of the system after appearance of the soft mode, we consider the following expression of magnetization called soft-mode-modulated magnetization

$$\mathbf{m}_a = (m + \tilde{m}_0 t)[-\sin(\theta + \tilde{\theta}_0 t)\sin(\varphi + \tilde{\psi}_0 t), \sin(\theta + \tilde{\theta}_0 t)\cos(\varphi + \tilde{\psi}_0 t), \cos(\theta + \tilde{\theta}_0 t)]^T, \quad (4)$$

where $\tilde{m}_0$, $\tilde{\theta}_0$, and $\tilde{\psi}_0$ denote unitized function of the vibration shape of the soft mode, and $t$ can be understood as the rescaled time or magnitude parameter. $\mathbf{m}_a(t)$ defined in Eq. (4) evolves with time, and we can calculate the corresponding emergent elastic coefficient $\tilde{C}^e_{\rho\rho}(\rho, t) = C^e_{\rho\rho}(\rho)\big|_{\mathbf{m}=\mathbf{m}_a(t)}$, which is called the soft-mode-modulated emergent elastic coefficient. We propose the following criterion for topological phase transitions with a soft mode:

*For a field solution with a soft mode, the occurrence of a topological phase transition initiates at some critical spatial point $\mathbf{x}_c$ and some critical soft mode magnitude $t_c$ where the positive-definiteness of the corresponding soft mode mediated emergent stiffness matrix $\tilde{\mathbf{C}}^e$ is broken.*

In the case of skyrmion annihilation phase transition at $b = b_a$, the only effective vibrational emergent elastic stiffness coefficient is $\tilde{C}^e_{\rho\rho}(\rho, t)$, and we show in Fig. 3(c) that the appearance of soft mode indeed leads to a softened point at $\rho_c = 0$. Since $C^e_{\rho\rho} = 2[m'^2(\rho) + m^2(\rho)\theta'^2(\rho)]$, we further know that this soft point stems from the vanishing of $m + \tilde{m}_0 t_c$ at $\rho_c = 0$ (Fig. 3(a)). In Fig. 4, we show the evolution of the magnetization pattern as $t$ increases from 0 to $t_c$.

Previous studies believe that annihilation of a skyrmion is initiated by a reversal of local magnetization inside the skyrmion, which require a sudden change of $\theta$ at some point by $\pi$. This explain why annihilation of a skyrmion is always a first-order phase transition. However, the reason for such a local reversal is thermal fluctuation, which occur randomly and leads people to wonder if the location of annihilation and its critical condition are exact and repeatable. From the above analysis, we show that annihilation of an isolated skyrmion always initiates at the center and the critical condition $b = b_a$ is exact. This is because the soft mode vanishes $m$ at and near by (concerning the boundary condition $\left(\frac{\partial m}{\partial \rho}\right)_{\rho=0} = 0$) the center, which means that $\theta$ can take arbitrary value at the center and merely change the free energy. In this case, reversal of local magnetization will definitely occur at the center because the local energy barrier for such a reversal approaches zero. In our study, annihilation of a 2D skyrmion is found to be induced by a 2D magnetic monopole at the center. In real materials with finite thickness, the translational symmetry in the z-axis is broken, and the skyrmion with a 3D structure is anticipated to be annihilated by appearence of a 3D magnetic monopole[6]. We believe the soft mode mechanism still works in that case, but its dynamics process is more complicated.

**Methods**

*A. Numerical method to obtain the isolated skyrmion solution*

The Euler-Lagrangian equation for $m(\rho)$ and $\theta(\rho)$ defined in eq. (2) can be derived as

$$Q_1(m,\theta) = m\left(\frac{\partial^2\theta}{\partial\rho^2} + \frac{1}{\rho}\frac{\partial\theta}{\partial\rho} - \frac{\sin 2\theta}{2\rho^2} + \frac{2\sin^2\theta}{\rho}\right) - b\sin\theta = 0, \quad (A1)$$

$$Q_2(m,\theta) = \frac{\partial^2 m}{\partial\rho^2} + \frac{1}{\rho}\frac{\partial m}{\partial\rho} - tm - 2m^3 + b\cos\theta - m\left(\left(\frac{\partial\theta}{\partial\rho}\right)^2 + \frac{\sin^2\theta}{\rho^2} + 2\frac{\partial\theta}{\partial\rho} + \frac{\sin 2\theta}{\rho}\right) = 0 \quad (A2)$$

with boundary conditions $\theta(0) = \pi$ and $\left(\frac{\partial m}{\partial\rho}\right)_{\rho=0} = 0$. If the radius of the skyrmion is $R$ (finite), we further have $\theta(R) = 0$.

We use the iteration method [] to numerically find the solution of $m(\rho)$ and $\theta(\rho)$ by considering the following decomposition

$$m(\rho) = m_a(\rho) + m_c(\rho), \quad (A3)$$

$$\theta(\rho) = \theta_a(\rho) + \theta_c(\rho), \quad (A4)$$

where $m_a(\rho)$ and $\theta_a(\rho)$ are two known functions we use to approach $m(\rho)$ and $\theta(\rho)$, and $m_c(\rho)$ and $\theta_c(\rho)$ are the corresponding correction function to be determined. Substituting eqs. (A3, A4) into eqs. (A1, A2) and linearize them with respect to $m_c(\rho)$ and $\theta_c(\rho)$, we have

$$L_{m1}m_c(\rho) + L_{\theta 1}\theta_c(\rho) = Q_1(m_a, \theta_a), \quad (A5)$$

$$L_{m2}m_c(\rho) + L_{\theta2}\theta_c(\rho) = Q_2(m_a, \theta_a), \tag{A6}$$

where $L_{mi}$ and $L_{\theta i}$, $(i = 1, 2)$ are the corresponding linear operators whose detailed expressions can be derived from eqs. (A1, A2) as

$$L_{m1} = -4mr(1 + \frac{\partial\theta}{\partial r})\frac{\partial}{\partial r} + 2m\left(-2 - 2\frac{\partial\theta}{\partial r} - 2r\frac{\partial^2\theta}{\partial r^2} + 2\cos 2\theta + \frac{\sin 2\theta}{r} + kr\sin 2\theta\right)$$
$$- 4\frac{\partial m}{\partial r}r(1 + \frac{\partial\theta}{\partial r}) + 2br\sin\theta,$$
$$\tag{A7}$$

$$L_{\theta1} = -2m^2r\frac{\partial^2}{\partial r^2} - (2m + 4r\frac{\partial m}{\partial r})m\frac{\partial}{\partial r} + m^2\left(\frac{2\cos 2\theta}{r} + 2kr\cos 2\theta - 4\sin 2\theta\right) + 2bmr\cos\theta,$$
$$\tag{A8}$$

$$L_{m2} = -2r\frac{\partial^2}{\partial r^2} - 2\frac{\partial}{\partial r} - kr + \frac{1}{r} + 4r\frac{\partial\theta}{\partial r} + 2r\left(\frac{\partial\theta}{\partial r}\right)^2 + 2tr - 6m^2tr - \left(kr + \frac{1}{r}\right)\cos 2\theta$$
$$+ 2\sin 2\theta,$$
$$\tag{A9}$$

$$L_{\theta2} = 4mr(1 + \frac{\partial\theta}{\partial r})\frac{\partial}{\partial r} + 2br\sin\theta + 4m\cos 2\theta + 2m(kr + \frac{1}{r})\sin 2\theta. \tag{A10}$$

Generally speaking, eqs. (A5, A6) cannot be solved analytically and finite difference method is applied to discretize them to a set of linear algebraic equations with solution

$$\mathbf{v} = \mathbf{L}^{-1}\mathbf{Q}, \tag{A6}$$

where

$$\mathbf{v} = [m_{c0}, m_{c1}, m_{c2}, \ldots m_{cn}, \theta_{c0}, \theta_{c1}, \theta_{c2}, \ldots \theta_{cn}]^T, \tag{A7}$$

$$\mathbf{Q} = [Q_{10}, Q_{11}, Q_{12}, \ldots Q_{1n}, Q_{20}, Q_{21}, Q_{22}, \ldots Q_{2n}]^T, \tag{A8}$$

$$\mathbf{L} = \begin{bmatrix} \mathbf{L}_{m1} & \mathbf{L}_{\theta1} \\ \mathbf{L}_{m2} & \mathbf{L}_{\theta2} \end{bmatrix}, \tag{A9}$$

and $m_c(\rho)$ and $\theta_c(\rho)$ are discretize into two vectors with $n + 1$ components denoted by $m_{ci}$ and $\theta_{ci}$, $(i = 0, 1, \ldots, n)$, and $\mathbf{L}_{mi}$ and $\mathbf{L}_{\theta i}$, $(i = 1, 2)$ are the discretized matrix form of the linear operator $L_{mi}$ and $L_{\theta i}$. In eq. (A6), $\mathbf{L}^{-1}$ denotes the inverse matrix of $\mathbf{L}$. By setting new approaching functions as follow

$$m_{a'}(\rho) = m_a(\rho) + m_c(\rho), \tag{A10}$$

$$\theta_{a'}(\rho) = \theta_a(\rho) + \theta_c(\rho), \tag{A11}$$

we repeat the process above to solve $m_{c'}(\rho)$ and $\theta_{c'}(\rho)$, and iterate until $m_{c'}(\rho)$ and $\theta_{c'}(\rho)$ reaches zero for all $\rho$. In this case the final version of $m_{a'}(\rho)$ and $\theta_{a'}(\rho)$ is the desired solution.

## B. Emergent elasticity and topological stability of a vector field in 3D space

For an arbitrary vector solution $\mathbf{p}(\mathbf{x}) = \tilde{\mathbf{p}}(\mathbf{x})$ with three components in 3D space, a general form of critical condition of topological stability can be derived following our previous work[19] and is presented as follow. For $\phi = \phi(\mathbf{x}, p_i, p_{j,k})$, we do the following replacements concerning domain emergent elasticity: $p_i(\mathbf{x}) \to \tilde{p}_i(\mathbf{x} - \mathbf{u}(\mathbf{x}))$, $p_{j,k}(\mathbf{x}) \to \tilde{p}_{j,i}(\mathbf{x} - \mathbf{u}(\mathbf{x}))(\delta_{ik} - u_{i,k}) = \tilde{p}_{j,i}(\mathbf{x} - \mathbf{u}(\mathbf{x}))(\delta_{ik} - (\varepsilon_{ik} + \epsilon_{ikl}\omega_l))$, and obtain $\phi_u = \phi(\mathbf{x}, \tilde{p}_i(\mathbf{x} - \mathbf{u}(\mathbf{x})), \tilde{p}_{j,i}(\mathbf{x} - \mathbf{u}(\mathbf{x}))(\delta_{ik} - (\varepsilon_{ik} + \epsilon_{ikl}\omega_l)))$, where $\varepsilon_{ik} = \frac{1}{2}(u_{i,k} + u_{k,i})$ is the domain emergent elastic strains and $\epsilon_{ikl}\omega_l = \frac{1}{2}(u_{i,k} - u_{k,i})$, with $\omega_l$ the emergent rotational angle and $\epsilon_{ikl}$ are components of the Levi-Civita tensor according to relevant definition in solid mechanics. In 3D space, we can define the following domain emergent elastic deformation vector

$$\boldsymbol{\varepsilon} = [\varepsilon_{11} \quad \varepsilon_{22} \quad \varepsilon_{33} \quad \varepsilon_{23} \quad \varepsilon_{13} \quad \varepsilon_{12} \quad \omega_1 \quad \omega_2 \quad \omega_3]^T, \tag{B1}$$

and the domain emergent elastic stiffness matrix $\mathbf{C}^\varepsilon$ is derived by

$$C_{ij}^\varepsilon = \left(\frac{\partial^2 \phi_u}{\partial \varepsilon_i \partial \varepsilon_j}\right)_{\mathbf{u}(\mathbf{x})=\mathbf{0}}. \tag{B2}$$

The critical condition of topological stability in this case is at some point or points in space, at least one of the eigenvalues of $\mathbf{C}^\varepsilon$ drops to zero. According to our previous study[19], in this case the range emergent elastic stiffness matrix is equivalent to $\mathbf{C}^\varepsilon$, so that $\mathbf{C}^\varepsilon$ can be referred to as the emergent elastic stiffness matrix.

For $\phi = \phi(\mathbf{x}, p_i)$, we do the following replacements concerning domain emergent elasticity: $p_i(\mathbf{x}) \to \tilde{p}_i(\mathbf{x} - \mathbf{u}(\mathbf{x}))$, and obtain $\phi_u = \phi(\mathbf{x}, \tilde{p}_i(\mathbf{x} - \mathbf{u}(\mathbf{x})))$, the state variable here is the emergent displacement vector

$$\mathbf{u} = [u_1 \quad u_2 \quad u_3]^T, \tag{B3}$$

and the the domain emergent displacement stiffness matrix $\mathbf{C}^u$ is derived by

$$C_{ij}^u = \left(\frac{\partial^2 \phi_u}{\partial u_i \partial u_j}\right)_{\mathbf{u}(\mathbf{x})=\mathbf{0}}. \tag{B4}$$

The critical condition of topological stability in this case is at some point or points in space, at least one of the eigenvalues of $\mathbf{C}^u$ drops to zero. For range emergent elasticity of this case, a condition similar to that of the second order variation of $\phi$ with respect to $p_i$ can be obtained, so that $\mathbf{C}^u$ can be referred to as the emergent displacement stiffness matrix.

Now we apply the general formula above to study the isolated magnetic skyrmion. The free energy density is given by eq. (1), which corresponds to the case $\phi = \phi(m_i, m_{j,k})$, and the soliton solution is given by eq. (2), which takes the form $\mathbf{m} = \mathbf{m}(\rho) = \mathbf{m}(\sqrt{x^2 + y^2})$. Since this is a 2D solution, eq. (B1) reduces to

$$\boldsymbol{\varepsilon} = [\varepsilon_{11} \quad \varepsilon_{22} \quad \varepsilon_{12} \quad \omega_3]^T. \tag{B5}$$

And for the free energy density given by eq. (1), to obtain the expression of $\phi_u$ we do the following replacements: $m_{1,1} \to (1-\varepsilon_{11})m_{1,1} - (\varepsilon_{12} - \omega_3)m_{1,2}$, $m_{1,2} \to -(\varepsilon_{12} + \omega_3)m_{1,1} + (1-\varepsilon_{22})m_{1,2}$, $m_{2,1} \to (1-\varepsilon_{11})m_{2,1} - (\varepsilon_{12} - \omega_3)m_{2,2}$, $m_{2,2} \to -(\varepsilon_{12} + \omega_3)m_{2,1} + (1-\varepsilon_{22})m_{2,2}$, $m_{3,1} \to (1-\varepsilon_{11})m_{3,1} - (\varepsilon_{12} - \omega_3)m_{3,2}$, $m_{3,2} \to -(\varepsilon_{12} + \omega_3)m_{3,1} + (1-\varepsilon_{22})m_{3,2}$, and from eq. (B2) the expressions of the emergent elastic stiffness matrix $\mathbf{C}^\varepsilon$ can be derived in the Cartesian coordinates. Since the solution of isolated magnetic skyrmion in this case is axially symmetric, it is more convenient to study its emergent elastic stiffness in polar coordinates. To achieve this, we recall the following coordinate transformation relation:

$$\hat{\boldsymbol{\varepsilon}} = \mathbf{K}\boldsymbol{\varepsilon}, \tag{B6}$$

where

$$\mathbf{K} = \begin{bmatrix} \cos^2\varphi & \sin^2\varphi & -\sin 2\varphi/2 & 0 \\ \sin^2\varphi & \cos^2\varphi & \sin 2\varphi/2 & 0 \\ \sin 2\varphi & -\sin 2\varphi & \cos 2\varphi & 0 \\ 0 & 0 & 0 & 1 \end{bmatrix}, \tag{B7}$$

$$\hat{\boldsymbol{\varepsilon}} = [\varepsilon_{\rho\rho} \quad \varepsilon_{\varphi\varphi} \quad \varepsilon_{\rho\varphi} \quad \omega_3]^T. \tag{B8}$$

The emergent elastic stiffness matrix presented in polar coordinates can be derived as

$$\hat{\mathbf{C}}^\varepsilon = \mathbf{K}^T \mathbf{C}^\varepsilon \mathbf{K}. \tag{B9}$$

When $\phi$ is given by eq. (1), we have after manipulation

$$\hat{\mathbf{C}}^\varepsilon = \begin{bmatrix} C^e_{\rho\rho} & 0 & 0 & 0 \\ 0 & C^e_{\varphi\varphi} & 0 & 0 \\ 0 & 0 & C^e_{\gamma\gamma} & C^{e\omega} \\ 0 & 0 & C^{e\omega} & C^{\omega_3} \end{bmatrix}, \tag{B10}$$

The four eigenvalues of $\hat{\mathbf{C}}^\varepsilon$ are $C^e_{\rho\rho} = 2[m'^2(\rho) + m^2(\rho)\theta'^2(\rho)]$, $C^e_{\varphi\varphi} = \frac{2m^2(\rho)\sin^2\theta(\rho)}{\rho^2}$, $C^e_3 = 4[m'^2(\rho) + m^2(\rho)\theta'^2(\rho)]$, and $C^e_4 = 4m^2(\rho)\frac{\sin^2\theta(\rho)}{\rho^2}$, and the corresponding eigenvectors are $[1 \ 0 \ 0 \ 0]^T$, $[0 \ 1 \ 0 \ 0]^T$, $[0 \ 0 \ 1 \ 1]^T$, and $[0 \ 0 \ -1 \ 1]^T$, respectively.

C. *Axially symmetric spin-wave modes of an isolated skyrmion*

Substituting eq. (4) into eq. (3), after manipulation we have

$$\ddot{\psi}_v = L_v^T \psi_v, \tag{C1}$$

$$\ddot{\theta}_v = L_v \theta_v, \tag{C2}$$

$$m_v = - L_{m2}^{-1} L_{\theta 2} \theta_v, \tag{C3}$$

where $L_v^T = -\left(\frac{M}{\gamma} m^3 \sin\theta\right)^{-1} L_\theta \left(\frac{M}{\gamma} m^3 \sin\theta\right)^{-1} L_\psi$, $L_v = -\left(\frac{M}{\gamma} m^3 \sin\theta\right)^{-1} L_\psi \left(\frac{M}{\gamma} m^3 \sin\theta\right)^{-1} L_\theta$, and

$$L_\psi = -2m^2 r \sin\theta^2 \frac{\partial^2}{\partial r^2} - 2m\left(2r\frac{\partial m}{\partial r}\sin\theta^2 + m(\sin\theta^2 + r\frac{\partial \theta}{\partial r}\sin 2\theta)\right)\frac{\partial}{\partial r}, \tag{C4}$$

$$L_\theta = - L_{m1} L_{m2}^{-1} L_{\theta 2} + L_{\theta 1}. \tag{C5}$$

The expressions of $L_{mi}$ and $L_{\theta i}$, $(i = 1, 2)$ are given in eqs. (A7-A10). Eq. (C1) and eq. (C2) correspond to the eigenvalue problem of the linear operator $L_v^T$ and $L_v$, which share the same eigenvalues but have different eigenvectors. $m_v$ is determined by $\theta_v$ through eq. (C3). Again we use finite difference method to discretize eq. (C1) and eq. (C2), which yields two eigenvalue value problems of matrices $\mathbf{L}_v^T$ and $\mathbf{L}_v$. $n$ axially symmetric spin-wave modes can be obtained if the domain $[0, \rho]$ is discretize into $n + 1$ points.

**Acknowledgments**

The work was supported by the NSFC (National Natural Science Foundation of China) through the funds with Grant Nos. 12172090, 11772360, 11832019, 11572355, the Natural Science Foundation of Guangdong Province (Grant No. 2019A1515012016), and Pearl River Nova Program of Guangzhou (Grant No. 201806010134).

**Author contributions**

Y.H. conceived the idea and conducted the work.

**Competing interests**

The authors declare no competing interests.

**References**

[1] Heisenberg, W. (1984). Introduction to the unified field theory of elementary particles. In *Scientific Review Papers, Talks, and Books Wissenschaftliche Übersichtsartikel, Vorträge und Bücher* (pp. 677-861). Springer, Berlin, Heidelberg.

[2] Skyrme, T. H. (1961). A non-linear field theory. *Proc. R. Soc. Lond. Ser. A 260*, 127–138.

[3] Weinberg, E. (2012). *Classical Solutions in Quantum Field Theory: Solitons and Instantons in High Energy Physics* (Cambridge Monographs on Mathematical Physics). Cambridge: Cambridge University Press. doi:10.1017/CBO9781139017787.

[4] Nguyen, J. H., Luo, D., & Hulet, R. G. (2017). Formation of matter-wave soliton trains by modulational instability. *Science, 356*(6336), 422-426.

[5] Carr, L. D., & Brand, J. (2004). Spontaneous soliton formation and modulational instability in Bose-Einstein condensates. *Physical Review Letters, 92*(4), 040401.

[6] Milde, P., Köhler, D., Seidel, J., Eng, L. M., Bauer, A., Chacon, A., ... & Rosch, A. (2013). Unwinding of a skyrmion lattice by magnetic monopoles. *Science, 340*(6136), 1076-1080.

[7] Qi, X. L., Li, R., Zang, J., & Zhang, S. C. (2009). Inducing a magnetic monopole with topological surface states. *Science, 323*(5918), 1184-1187.

[8] Scott, A. (1992). Davydov's soliton. *Physics Reports, 217*(1), 1-67.

[9] Su, W., Schrieffer, J. R., & Heeger, A. J. (1979). Solitons in polyacetylene. *Physical Review Letters, 42*(25), 1698.

[10] Khaykovich, L., Schreck, F., Ferrari, G., Bourdel, T., Cubizolles, J., Carr, L. D., ... & Salomon, C. (2002). Formation of a matter-wave bright soliton. *Science, 296*(5571), 1290-1293.

[11] Fu, Q., Wang, P., Huang, C., Kartashov, Y. V., Torner, L., Konotop, V. V., & Ye, F. (2020). Optical soliton formation controlled by angle twisting in photonic moiré lattices. *Nature Photonics, 14*(11), 663-668.

[12] Lee, T. D. (1987). Soliton stars and the critical masses of black holes. *Physical Review D, 35*(12), 3637.

[13] Eto, M., Isozumi, Y., Nitta, M., Ohashi, K., & Sakai, N. (2005). Instantons in the Higgs phase. *Physical Review D, 72*(2), 025011.

[14] Manton, N., & Sutcliffe, P. (2004). *Topological Solitons* (Cambridge Monographs on

*Mathematical Physics). Cambridge: Cambridge University Press. doi:10.1017/CBO9780511617034*

[15] *Brown, G. E., & Rho, M. (2010). The multifaceted skyrmion. World Scientific.*

[16] *Chaikin, P. M., Lubensky, T. C., & Witten, T. A. (1995). Principles of condensed matter physics (Vol. 10). Cambridge: Cambridge university press.*

[17] *Hu, Y., (2022). General theory of emergent elasticity for second-order topological phase transitions . Submitted.*

[18] *Skyrme, T. H. (1962). A unified field theory of mesons and baryons. Nuclear Physics, 31, 556-569.*

[19] *Leslie, L. S., Hansen, A., Wright, K. C., Deutsch, B. M. & Bigelow, N. P. (2009). Creation and detection of skyrmions in a bose-einstein condensate. Phys. Rev. Lett. 103, 250401, https://doi.org/10.1103/PhysRevLett.103.250401.*

[20] *Bychkov, Yu. A. and Maniv, T. and Vagner, I. D. (1996). Charged Skyrmions: A condensate of spin excitons in a two-dimensional electron gas. Phys. Rev. B 53, 10148.*

[21] *Fukuda, J. and Žumer, S. (2011). Quasi-two-dimensional Skyrmion lattices in a chiral nematic liquid crystal. Nat. Commun. 2:246 doi: 10.1038/ncomms1250.*

[22] *Bogdanov, A. and Hubert, A.(1994). The Properties of Isolated Magnetic Vortices. Journal of magnetism and magnetic materials, 186(2), 527–543. doi:10.1002/pssb.2221860223.*

[23] *Mühlbauer, S., Binz, B., Jonietz, F., Pfleiderer, C., Rosch, A., Neubauer, A., ... & Böni, P. (2009). Skyrmion lattice in a chiral magnet. Science, 323(5916), 915-919.*

[24] *Du, L., Yang, A., Zayats, A.V. et al. (2019). Deep-subwavelength features of photonic skyrmions in a confined electromagnetic field with orbital angular momentum. Nat. Phys. 15, 650–654 https://doi.org/10.1038/s41567-019-0487-7.*

[25] *Roessler, U. K., Bogdanov, A. N., & Pfleiderer, C. (2006). Spontaneous skyrmion ground states in magnetic metals. Nature, 442(7104), 797-801.*

[26] *Liu, Y., Yin, G., Zang, J.D., Shi, J., & Lake, R.K. (2015). Skyrmion creation and annihilation by spin waves. Appl. Phys. Lett. 107, 152411.*

[27] *Desplat, L., Kim, J.-V. & Stamps, R. L. (2019). Paths to annihilation of first- and second-order (anti)skyrmions via (anti)meron nucleation on the frustrated square lattice. Phys. Rev. B, 99, 174409.*

[28] Birch, M.T., Cortés-Ortuño, D., Khanh, N.D. et al. (2021). Topological defect-mediated skyrmion annihilation in three dimensions. Commun. Phys. 4, 175.

[29] Kathinka, G., Bastian, P., Felix, B. et al. (2021). Application concepts for ultrafast laser-induced skyrmion creation and annihilation. Appl. Phys. Lett. 118, 192403.

[30] Feng, C., Meng, F., Wang, Y., et al., (2021). Field-Free Manipulation of Skyrmion Creation and Annihilation by Tunable Strain Engineering. Adv. Funct. Mater. 31, 2008715.

[31] Li, Y., & Pang, H. (2022). Annihilation dynamics of magnetic skyrmion lattice state under in-plane magnetic field and spontaneous recovery after field removal. Phys. Rev. B 105, 174425.

[32] Fert, A., Reyren, N., & Cros, V. (2017). Magnetic skyrmions: advances in physics and potential applications. Nat. Rev. Mater. 2, 17031.

[33] Fert, A., Cros, V. & Sampaio, J. (2013). Skyrmions on the track. Nature Nanotech 8, 152–156

[34] Grollier, J., Querlioz, D., Camsari, K.Y. et al. (2020). Neuromorphic spintronics. Nat Electron 3, 360–370

[35] Song, K.M., Jeong, JS., Pan, B. et al. (2020). Skyrmion-based artificial synapses for neuromorphic computing. Nat Electron 3, 148–155

[36] Zázvorka, J., Jakobs, F., Heinze, D. et al. (2019). Thermal skyrmion diffusion used in a reshuffler device. Nat. Nanotechnol. 14, 658–661

[37] Leonov, A. O., & Bogdanov, A. N. (2018). Crossover of skyrmion and helical modulations in noncentrosymmetric ferromagnets. New Journal of Physics, 20(4), 043017.

[38] Kosterlitz, J. M. and Thouless, D. J. (1972). Long range order and metastability in two-dimensional solids and superfluids. J. Phys. C: Solid State Phys. 5 L124–6.

[39] Kosterlitz, J. M. and Thouless, D. J. (1973). Ordering, metastability and phase transitions in two-dimensional systems. J. Phys. C: Solid State Phys. 6 1181–203.

[40] Landau, L. D. & Lifshitz, E. M. Theory of the dispersion of magnetic permeability in ferromagnetic bodies. Phys. Z. Sowietunion. 8 153169 (1935).

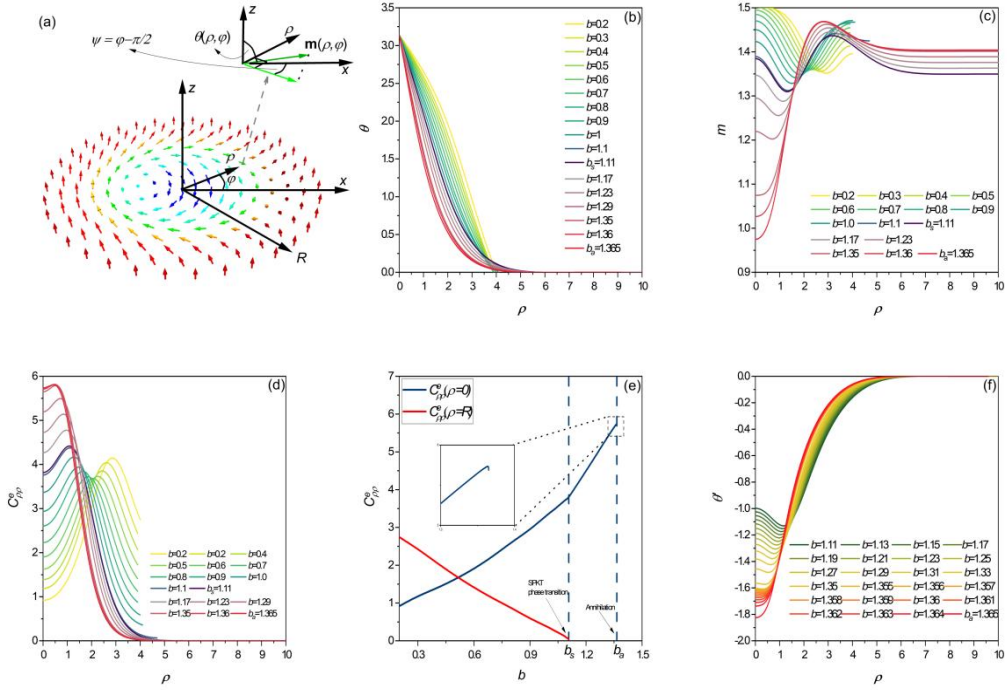

Figure 1. (a) Illustration of the field pattern of a Bloch-type isolated magnetic skyrmion. Variation in (b) $\theta$, (c) $m$, (d) $C^e_{\rho\rho}$ and (f) $\theta'$, and (d) $C^e_{\rho\rho}$ with $\rho$ at different values of $b$. (d) Variation in the minimized free energy density $\phi$ with the radius of an isolated skyrmion $R$ calculated at different values of $b$. The circular blue dots mark the minimum point of the $\phi - R$ curves calculated at different $b$, while the blue star marks the appearance of a local maximum of the $\phi - R$ curve calculated at $b = 1.11$, and $\phi$ decreases with $R$ after this point such that the equilibrium $R$ approaches infinity at $b = 1.11$. (f) Variation in $C^e_{\rho\rho}(\rho = 0)$ and $C^e_{\rho\rho}(\rho = R)$ with the applied magnetic field $b$.

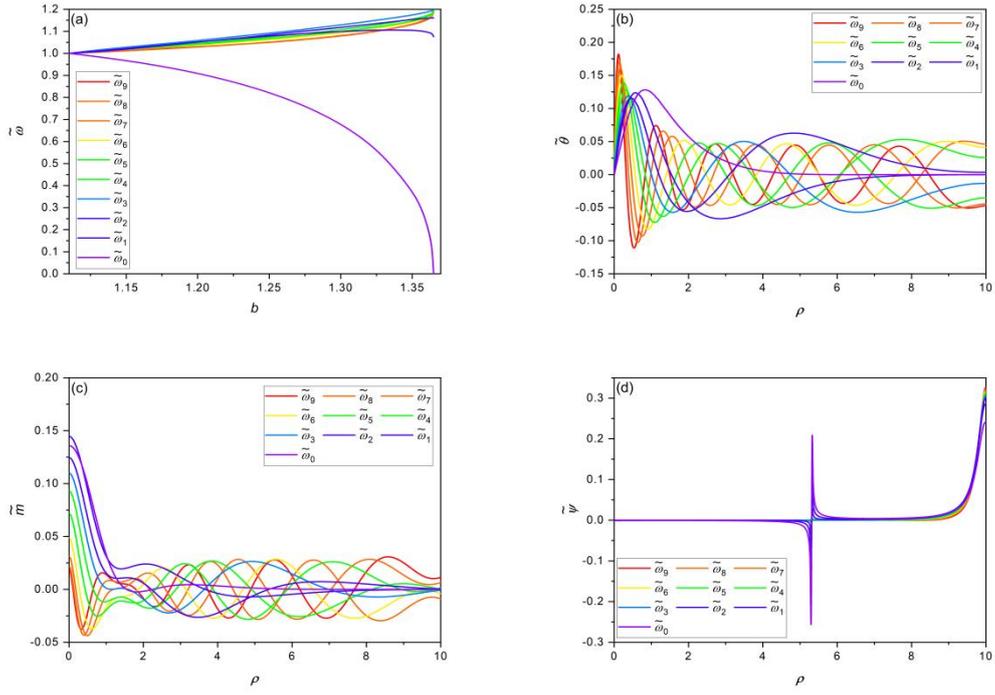

Figure 2. (a) Rescaled magnon frequencies, and the corresponding vibrational shape of (b) $\tilde{\theta}(\rho)$, (c) $\tilde{m}(\rho)$, and (d) $\tilde{\psi}(\rho)$ for the first ten spin-wave modes $\tilde{\omega}_i(b) = \omega_i(b)/\omega_i(b = 1.11)$, ($i = 0, 1, 2, \ldots, 9$). At $b = b_a$, $\tilde{\omega}_0$ vanishes and become a soft mode. The actually values of $\omega_i(b = 1,1)$ are listed as follow: $\omega_0^2(b = 1.11) = 3.74948$, $\omega_1^2(b = 1.11) = 33.17803$, $\omega_2^2(b = 1.11) = 71.0732$, $\omega_3^2(b = 1.11) = 119.22972$, $\omega_4^2(b = 1.11) = 193.81295$, $\omega_5^2(b = 1.11) = 307.57504$, $\omega_6^2(b = 1.11) = 481.11104$, $\omega_7^2(b = 1.11) = 740.41024$, $\omega_8^2(b = 1.11) = 1594.75047$, $\omega_9^2(b = 1.11) = 1101.3593$.

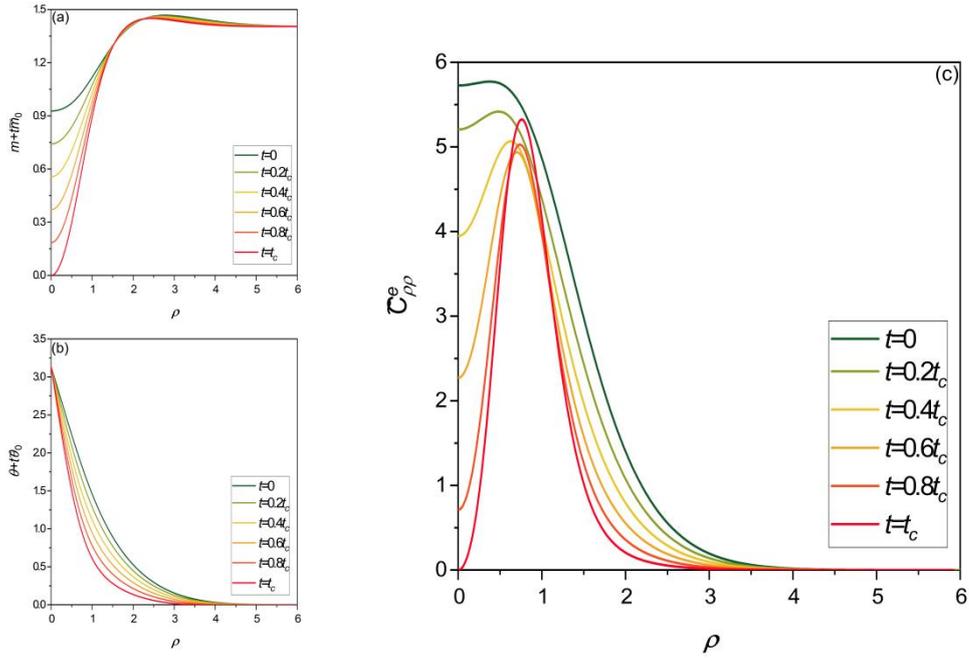

Figure 3. Variation in (a) $m + t\tilde{m}_0$, (b) $\theta + t\tilde{\theta}_0$, and (c) $\tilde{C}^e_{\rho\rho}$ with $\rho$ at different values of $t$.

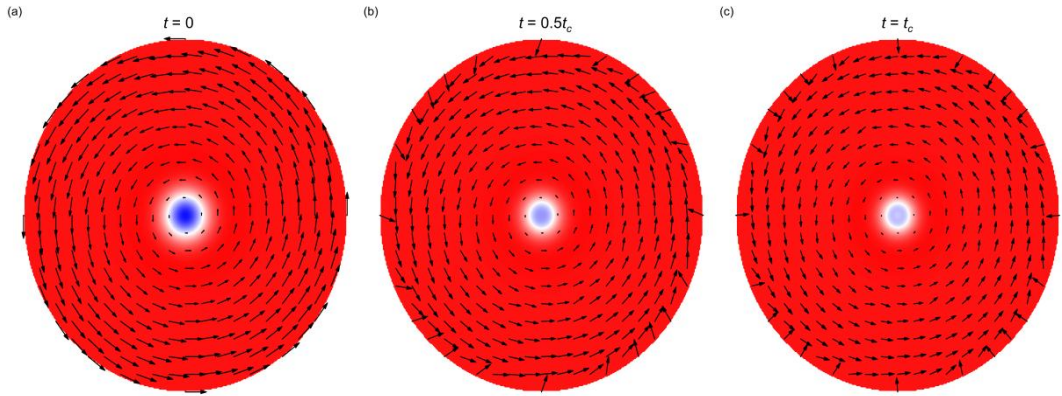

Figure 4. Field configuration of the soft-mode-modulated magnetization at (a) $t = 0$, (b) $t = 0.5t_c$, and (c) $t = t_c$, where the vectors illustrate the distribution of the in-plane magnetization components with length proportional to their magnitude, and the colored density plot illustrates the distribution of the out-of-plane magnetization component (red means pointing upward and blue means pointing downward).